\documentstyle[11pt,emulateapj,times]{article}
 
\newcommand{\etal}{et~al.\ }
\newcommand{\PVdblt}{{\rm P}\kern 0.1em{\sc v}~$\lambda\lambda 1117, 1128$}
\newcommand{\CIVdblt}{{\rm C}\kern 0.1em{\sc iv}~$\lambda\lambda 1548, 1550$}
\newcommand{\MgIIdblt}{{\rm Mg}\kern 0.1em{\sc ii}~$\lambda\lambda 2796, 2803$}
\newcommand{\NVdblt}{{\rm N}\kern 0.1em{\sc v}~$\lambda\lambda 1238, 1242$}  
\newcommand{\SVIdblt}{{\rm S}\kern 0.1em{\sc vi}~$\lambda\lambda 933, 944$} 
\newcommand{\OVIdblt}{{\rm O}\kern 0.1em{\sc vi}~$\lambda\lambda 1031, 1037$} 
\newcommand{\SiIIdblt}{{\rm Si}\kern 0.1em{\sc ii}~$\lambda\lambda 1190, 1193$} 
\newcommand{\SiIVdblt}{{\rm Si}\kern 0.1em{\sc iv}~$\lambda\lambda 1393, 1402$} 
\newcommand{\PV}{\hbox{{\rm P}\kern 0.1em{\sc v}}}
\newcommand{\AlI}{\hbox{{\rm Al}\kern 0.1em{\sc i}}}
\newcommand{\AlII}{\hbox{{\rm Al}\kern 0.1em{\sc ii}}}
\newcommand{\AlIII}{{\hbox{\rm Al}\kern 0.1em{\sc iii}}}
\newcommand{\CaII}{\hbox{{\rm Ca}\kern 0.1em{\sc ii}}}
\newcommand{\CII}{\hbox{{\rm C}\kern 0.1em{\sc ii}}}
\newcommand{\CIIe}{\hbox{{\rm C$^{\ast}$}\kern 0.1em{\sc ii}}}
\newcommand{\CIII}{\hbox{{\rm C}\kern 0.1em{\sc iii}}}
\newcommand{\CIV}{\hbox{{\rm C}\kern 0.1em{\sc iv}}}
\newcommand{\CV}{\hbox{{\rm C}\kern 0.1em{\sc v}}}
\newcommand{\HI}{\hbox{{\rm H}\kern 0.1em{\sc i}}}
\newcommand{\HII}{\hbox{{\rm H}\kern 0.1em{\sc ii}}}
\newcommand{\Lya}{\hbox{{\rm Ly}\kern 0.1em$\alpha$}}
\newcommand{\Lyb}{\hbox{{\rm Ly}\kern 0.1em$\beta$}}
\newcommand{\Lyg}{\hbox{{\rm Ly}\kern 0.1em$\gamma$}}
\newcommand{\Lyd}{\hbox{{\rm Ly}\kern 0.1em$\delta$}}
\newcommand{\Lye}{\hbox{{\rm Ly}\kern 0.1em$\epsilon$}}
\newcommand{\Lyphi}{\hbox{{\rm Ly}\kern 0.1em$\phi$}}
\newcommand{\Lyfive}{\hbox{{\rm Ly}\kern 0.1em$5$}}
\newcommand{\Lysix}{\hbox{{\rm Ly}\kern 0.1em$6$}}
\newcommand{\Lyseven}{\hbox{{\rm Ly}\kern 0.1em$7$}}
\newcommand{\Lyeight}{\hbox{{\rm Ly}\kern 0.1em$8$}}
\newcommand{\Lynine}{\hbox{{\rm Ly}\kern 0.1em$9$}}
\newcommand{\Lyten}{\hbox{{\rm Ly}\kern 0.1em$10$}}
\newcommand{\HeI}{\hbox{{\rm He}\kern 0.1em{\sc i}}}
\newcommand{\HeII}{\hbox{{\rm He}\kern 0.1em{\sc ii}}}
\newcommand{\FeI}{\hbox{{\rm Fe}\kern 0.1em{\sc i}}}
\newcommand{\FeII}{\hbox{{\rm Fe}\kern 0.1em{\sc ii}}}
\newcommand{\FeIII}{\hbox{{\rm Fe}\kern 0.1em{\sc iii}}}
\newcommand{\MnII}{\hbox{{\rm Mn}\kern 0.1em{\sc ii}}}
\newcommand{\MgI}{\hbox{{\rm Mg}\kern 0.1em{\sc i}}}
\newcommand{\MgII}{\hbox{{\rm Mg}\kern 0.1em{\sc ii}}}
\newcommand{\MgIII}{\hbox{{\rm Mg}\kern 0.1em{\sc iii}}}
\newcommand{\NI}{\hbox{{\rm N}\kern 0.1em{\sc i}}}
\newcommand{\NII}{\hbox{{\rm N}\kern 0.1em{\sc ii}}}
\newcommand{\NIII}{\hbox{{\rm N}\kern 0.1em{\sc iii}}}
\newcommand{\NV}{\hbox{{\rm N}\kern 0.1em{\sc v}}}
\newcommand{\OVI}{\hbox{{\rm O}\kern 0.1em{\sc vi}}}
\newcommand{\OI}{\hbox{{\rm O}\kern 0.1em{\sc i}}}
\newcommand{\OII}{\hbox{[{\rm O}\kern 0.1em{\sc ii}]}}
\newcommand{\OIV}{\hbox{{\rm O}\kern 0.1em{\sc iv}]}}
\newcommand{\SVI}{{\rm S}\kern 0.1em{\sc vi}}
\newcommand{\SiI}{\hbox{{\rm Si}\kern 0.1em{\sc i}}}
\newcommand{\SiII}{\hbox{{\rm Si}\kern 0.1em{\sc ii}}}
\newcommand{\SiIII}{\hbox{{\rm Si}\kern 0.1em{\sc iii}}}
\newcommand{\SiIV}{\hbox{{\rm Si}\kern 0.1em{\sc iv}}}
\newcommand{\SII}{\hbox{{\rm S}\kern 0.1em{\sc ii}}}
\newcommand{\SIII}{\hbox{{\rm S}\kern 0.1em{\sc iii}}}
\newcommand{\NaI}{\hbox{{\rm Na}\kern 0.1em{\sc i}}}
\newcommand{\kms}{\hbox{km~s$^{-1}$}}

\tighten
\begin{document}
 
 
\lefthead{CHURCHILL ET~AL. }
\righthead{{\CIV}--{\MgII} CONNECTION IN GALAXIES}
 
\submitted{Astrophysical Journal Letters, {\it accepted}}
 
\title{\vglue -0.5in
The {\CIV} Absorption--{\MgII} Kinematics Connection in
$\left< \lowercase{z}\right> \sim 0.7$ Galaxies}

\thispagestyle{empty}
 
\author{\vglue -0.3in
Christopher~W.~Churchill\altaffilmark{1}, 
Richard~R.~Mellon\altaffilmark{1}, 
Jane~C.~Charlton\altaffilmark{1}, 
Buell~T.~Jannuzi\altaffilmark{2},
Sofia~Kirhakos\altaffilmark{3},
Charles~C.~Steidel\altaffilmark{4}, 
and
Donald~P.~Schneider\altaffilmark{1}
}

\affil{$^{1}$ The Pennsylvania State University, University Park, PA
16802, cwc, rmellon, charlton, dps@astro.psu.edu \\
$^{2}$ National Optical Astronomy Observatories, Tucson, AZ 85719,
jannuzi@noao.edu \\
$^{3}$ Institute for Advanced Study, Princeton, NJ 08544,
sofia@sns.ias.edu \\
$^{4}$ California Institute of Technology, Pasadena, CA 91125,
ccs@astro.caltech.edu }

\begin{abstract}
We have examined Faint Object Spectrograph data from the {\it Hubble
Space Telescope\/} Archive for {\CIVdblt} absorption associated with
$40$ {\MgIIdblt} absorption--selected galaxies at $0.4 \leq z \leq
1.4$. 
We report a strong correlation between {\MgII} kinematics, measured 
in $\sim 6$~{\kms} resolution HIRES/Keck spectra, and $W_{r}(1548)$;
this implies a physical connection between the processes that produce
``outlying velocity'' {\MgII} clouds and high ionization galactic/halo gas.
We found no trend in ionization condition, $W_{r}(1548)/W_{r}(2796)$,
with galaxy--QSO line--of--sight separation for 13 systems with confirmed
associated galaxies, suggesting no obvious ionization gradient with
galactocentric distance in these higher redshift galaxies.
We find tentative evidence ($2\sigma$) that $W_{r}(1548)/W_{r}(2796)$
is anti--correlated with galaxy $B-K$ color; if further data
corroborate this trend, in view of the strong {\CIV}--{\MgII}
kinematics correlation, it could imply a connection between stellar
populations, star formation episodes, and the kinematics and
ionization conditions of halo gas at $z\sim 1$.
\end{abstract}

\keywords{quasars--- absorption lines; galaxies--- evolution;
galaxies--- halos}

\section{Introduction}
\label{sec:intro}
 
The central and complex role of galactic gas in the star formation,
dynamical, and chemical evolution of galaxies is well established.
Evidence is mounting that, at the present epoch, multiphase gaseous
halos are a physical extension of their host galaxy's interstellar
medium (ISM); their physical extent, spatial distribution, ionization
conditions, and chemical enrichment are intimately linked to the
energy density rate infused into the galaxy's ISM by stellar winds and
ionizing radiation, and by supernovae shock waves (e.g.\
\cite{dahlem98}, and references therein). 
One long--standing question is how the halos and ISM of earlier epoch
galaxies compare or relate, in an evolutionary sense, to those of the
present epoch.

Normal ($\sim L^{\ast}$) galaxies at intermediate redshifts ($0.5 \leq
z \leq 1.0$) are seen to give rise to low ionization
{\MgIIdblt} absorption with $W_{r}(2796) \geq 0.3$~{\AA} out to
projected distances of $\sim 40h^{-1}$~kpc (e.g.\ \cite{steidel95}).
A key question is whether low ionization gas at large
galactocentric distances is due to infall (i.e.\ satellite accretion,
minor mergers, intragroup or intergalactic infall), or to energetic
processes in the ISM (galactic fountains, chimneys).
Using high resolution {\MgII} profiles, Churchill, Steidel, \& Vogt
(1996\nocite{csv96}) found no suggestive trends between low ionization
gas properties and galaxy properties at $\left< z \right> = 0.7$.
A next logical step toward addressing this question is to explore the
high ionization gas in {\MgII} absorption selected galaxies at these
redshifts.

The {\CIVdblt} doublet is a sensitive probe of higher ionization gas.
Using {\CIV} and {\MgII}, Bergeron \etal (1994\nocite{bergeron94})
inferred multiphase ionization structure around a $z=0.79$ galaxy.
Churchill \& Charlton (1999\nocite{q1206}) incorporated the {\MgII}
kinematics and found high metallicity, multiphase absorption in 
a possible group of three galaxies at $z=0.93$.
In a survey of the 3C~336 field (Q~$1622+238$), Steidel \etal
(1997\nocite{3c336paper}) reported that $W_{r}(1548)/W_{r}(2796)$
appeared to be correlated with galaxy--QSO impact parameter, as
expected if halo gas density decreases with galactocentric distance.

In this {\it Letter}, we present a study of {\CIV} absorption 
associated with $0.4 \leq z \leq 1.4$ {\MgII} absorption--selected 
galaxies using Faint Object Spectrograph (FOS) data available 
in the {\it Hubble Space Telescope\/} ({\it HST}) Archive. 
We compare the {\CIV} strengths to the {\MgII} strengths and
kinematics and (when available) to the galaxy properties.
 
\begin{figure*}[t]
\plotfiddle{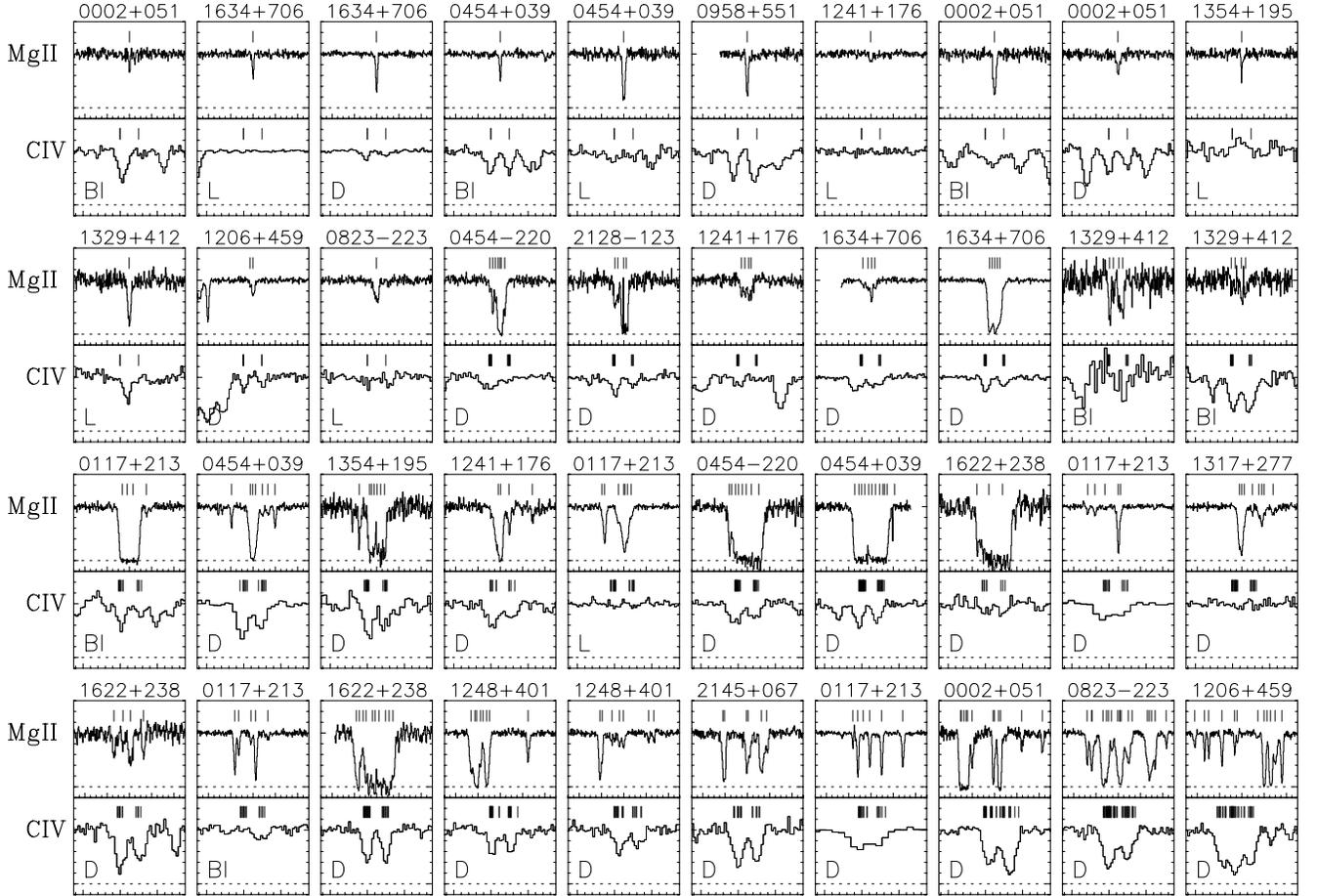}{4.7in}{0.}{78.}{78.}{-306}{-80}
\vglue -0.1in
\protect\caption
{\footnotesize
The normalized spectroscopic data with the zero point given by the
horizontal dotted line.  The {\MgII} velocity scale is $500$~{\kms}
and the {\CIV} velocity scale is $3000$~{\kms}.
--- (upper sub--panels) The HIRES {\MgII} $\lambda 2796$ transition is
plotted in order of kinematic spread, from upper left to lower right. 
--- (lower sub--panels) The corresponding FOS/{\it HST\/} spectrum is
plotted with ticks marking the expected location of the {\CIVdblt} doublet,
based upon Voigt profile fits to {\MgII}.  The labels ``D'', ``L'',
and ``Bl'' denote detection, limit, and blend, respectively.
\label{fig:fig1}
\vglue -0.2in
}
\end{figure*}

\section{The Data}
\label{sec:data}

The {\MgII} absorbers are selected from the HIRES/Keck sample of
Churchill (1997\nocite{thesis}) and Churchill \etal
(1999a\nocite{weak}). 
The HIRES resolution is $\sim 6$~{\kms} (\cite{vogt94}).
The data were processed using IRAF\footnote{IRAF is distributed by the
National Optical Astronomy Observat-- \\ ories, which are operated by
AURA, Inc., under contract to the NSF.}, as described in Churchill
\etal (1999a\nocite{weak}).
The redshifts of the individual {\MgII} sub--components were obtained
using {\sc minfit} (\cite{thesis}), a Voigt profile (VP) fitter that
uses $\chi ^{2}$ minimization.
For $36$ of the {\MgII} absorbers, FOS/{\it HST\/} (resolution $\sim
230$~{\kms}) spectra covering {\CIV} were available from the {\it
HST\/} Archive.
For four absorbers, {\CIV} was taken from ground--based
spectra of Steidel \& Sargent (1992\nocite{ss92}), and Sargent,
Boksenberg, \& Steidel (1988\nocite{sbs88}).
The FOS spectra were processed using the
techniques of the {\it HST\/} QSO Absorption Line Key
Project (\cite{dpsKP}; \cite{KP13}).
For 13 systems, the absorbing galaxy impact parameters, rest--frame
$K$ and $B$ luminosities, and $B-K$ colors are available from Steidel,
Dickinson, \& Persson (1994\nocite{sdp94}), Churchill \etal
(1996\nocite{csv96}), and Steidel \etal (1997\nocite{3c336paper}).
We will present a more detailed account in a companion paper
(\cite{archive}).


\section{Results}
\label{sec:results}

In Figure~\ref{fig:fig1}, we present the {\MgII} and {\CIV} data for
each of the 40 systems (note that the velocity scale for {\MgII} is
500~{\kms} and for {\CIV} is 3000~{\kms}).
Ticks above the HIRES spectra give the velocities of the multiple
VP {\MgII} sub--components and ticks above the FOS data give the
expected location of these components for both members of the {\CIV}
doublet.
The {\MgII} profiles are shown in order of increasing kinematic spread
from the upper left to lower right.
The kinematic spread is the second velocity moment of the
apparent optical depth of the {\MgII} $\lambda 2796$ profile, given by
$ \omega _{v}^{2} = \int \tau _{a}(v) v^{2}dv /
\int \tau _{a}(v)dv$, where $\tau _{a}(v) = \ln [I_{c}(v)/I(v)]$, and
$I(v)$ and $I_{c}(v)$ are the observed flux and the fitted continuum
flux at velocity $v$, respectively.
The zero--point velocity is given by the optical depth median of the
{\MgII} $\lambda 2796$ profile.
The kinematic spreads range from a few to $\sim 120$~{\kms}.

\begin{figure*}[b]
\plotfiddle{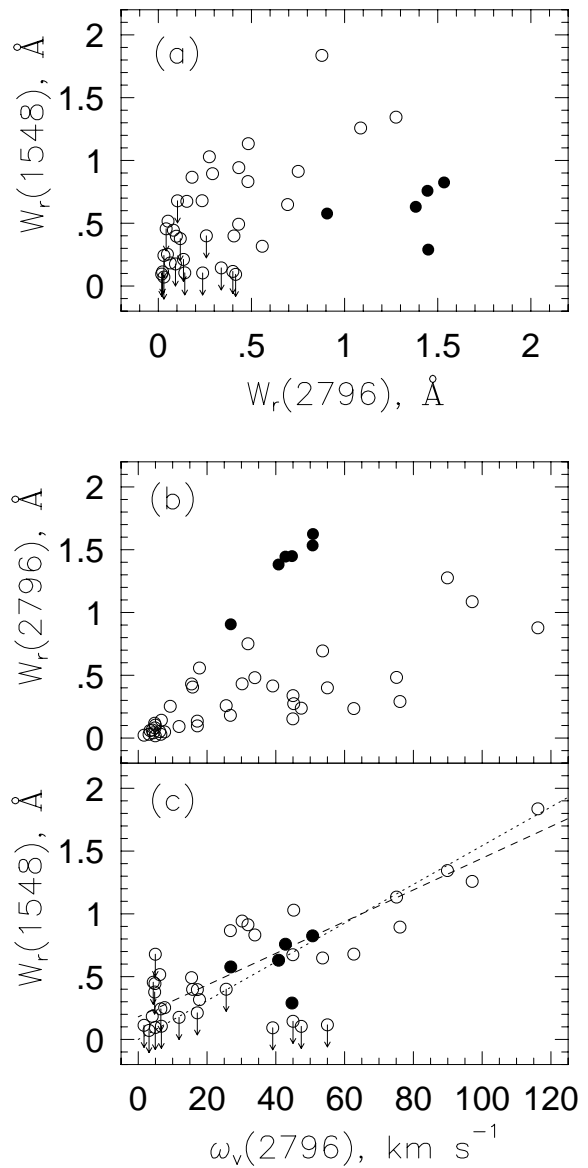}{0.in}{0.}{65.}{65.}{264}{60}
\parbox{3.5in}{\phantom{dummy}} \hfill
\parbox{3.5in}{\protect\caption{\vglue -1.8in 
\footnotesize
{\sc Fig}.~~2.
--- (a) $W_{r}(2796)$ vs.\ $W_{r}(1548)$.
--- (b) $W_{r}(2796)$ vs.\ the {\MgII} kinematic
spread, $\omega _{v}$.
--- (c) $W_{r}(1548)$ vs.\ $\omega _{v}$.  
Errors are on the order of the data point sizes.
Filled points are DLAs and candidate DLAs. The dotted and dashed
curves (panel c) are from linear fits, excluding
the upper limits (see text).
\label{fig:fig2}}}
\end{figure*}

In Figure~\ref{fig:fig2}$a$, we present $W_{r}(1548)$ vs.\
$W_{r}(2796)$, which exhibits considerable scatter.
Solid data points are damped {\Lya} absorbers (DLAs) and
candidate DLAs, based upon $W_{r}({\Lya}) \geq 8$~{\AA} or 
$W_{r}({\MgII}~\lambda 2796) \simeq W_{r}({\FeII}~\lambda
2600) \ga 1.0$~{\AA} (\cite{boisse}).

\begin{figure*}[t]
\vglue -0.1in
\plotfiddle{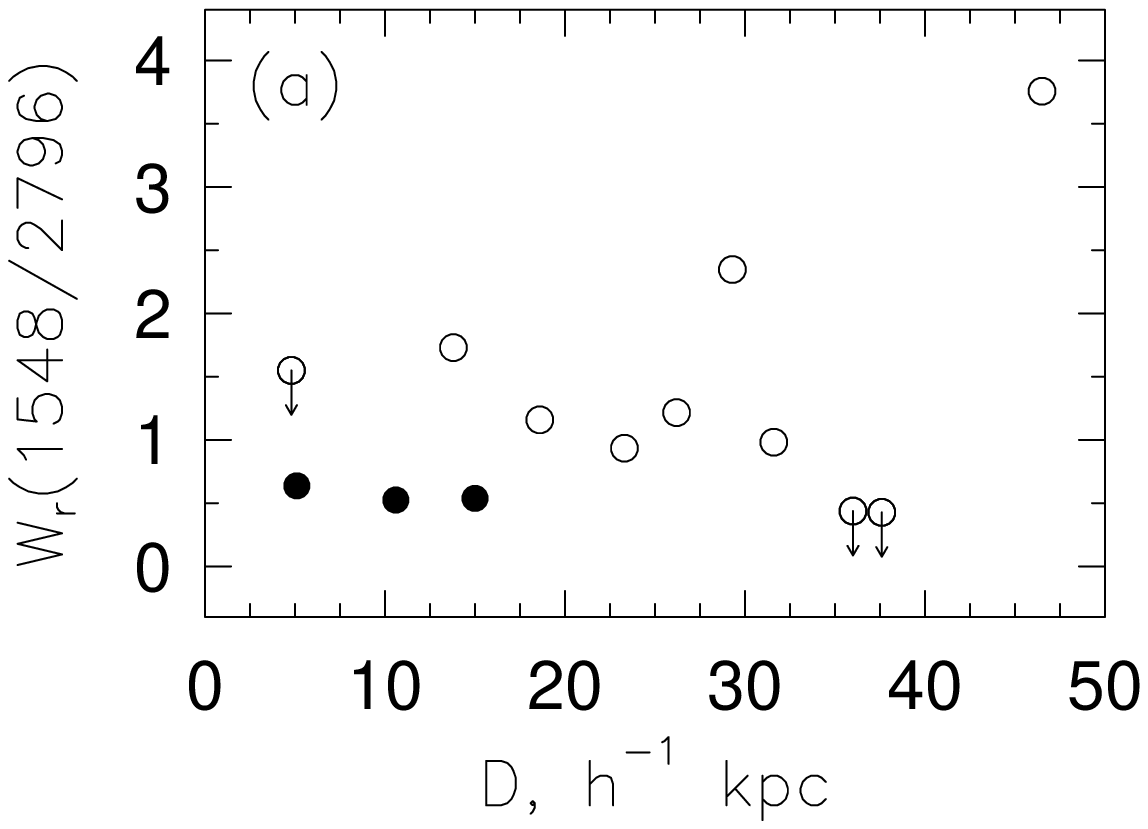}{1.9in}{0.}{75.}{75.}{-289}{-69}
\plotfiddle{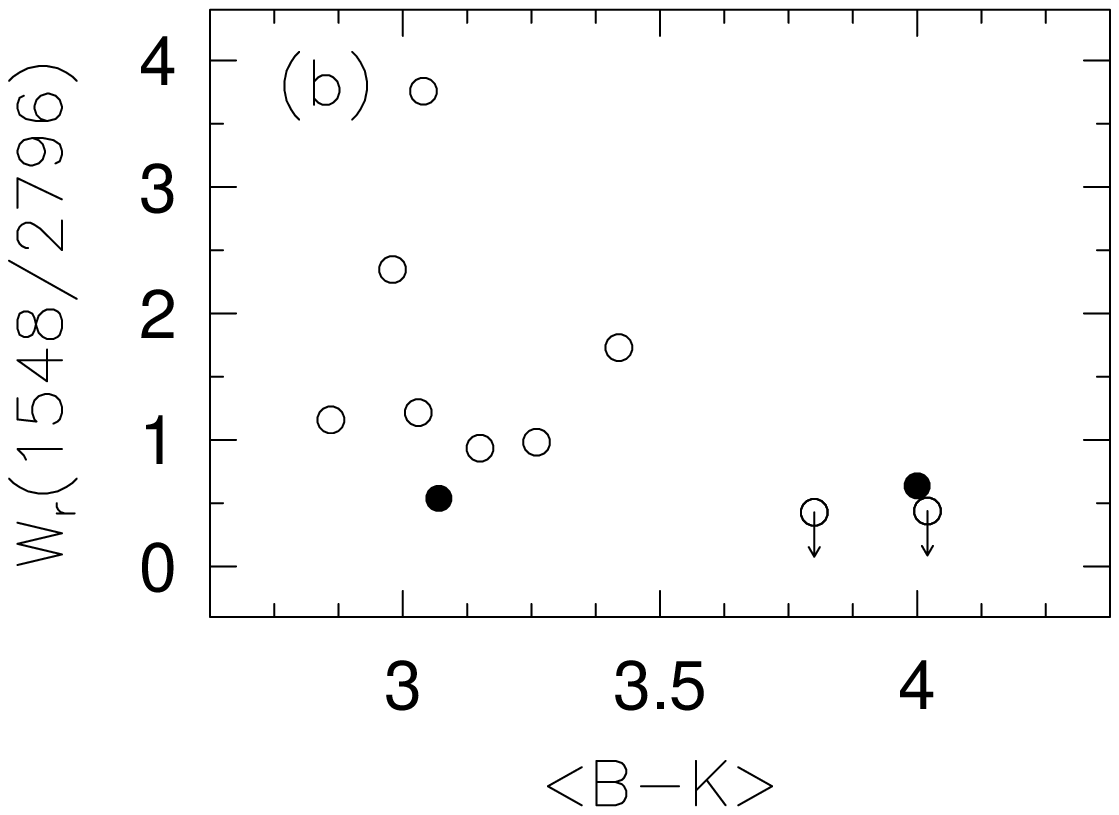}{0.0in}{0.}{75.}{75.}{-26}{-44}
\protect\caption{
\footnotesize
(a) $W_{r}(1548)/W_{r}(2796)$ vs.\ galaxy impact parameter, $D$.
--- (b) $W_{r}(1548)/W_{r}(2796)$ vs.\ galaxy rest--frame color,
$B-K$.
Errors are on the order of the data point sizes.
\label{fig:fig3}}
\vglue -0.2in
\end{figure*}

As seen in Figure~\ref{fig:fig2}$b$, $W_{r}(2796)$ correlates with
$\omega _{v}$.
Significant scatter arises because $\omega _{v}$ is sensitive
to the line--of--sight chance presence and equivalent width
distribution of the smaller $W_{r} (2796)$, ``outlying velocity'' clouds
(see \cite{kinematicpaper}).
The DLAs define a ``saturation line''; profiles with $W_{r}(2796) >
0.3$~{\AA} along this line have saturated cores.
As seen in Figure~\ref{fig:fig2}$c$, there is a tight correlation
between $\omega _{v}$ and $W_{r}(1548)$.
A Spearman--Kendall test, incorporating limits (\cite{eric}), yielded a
greater than  99.99\% confidence level.
A weighted least--squares fit to the data (dotted line through the
origin and with upper limits excluded) yielded a slope of $\omega _{v}
\simeq 65$~{\kms} per $1$~{\AA} of $W_{r}(1548)$.
The data exhibit a scatter of $\sigma _{W_{r}}(1548) = 0.22$~{\AA}
about the fit.
An essentially identical  maximum likelihood fit is shown as a dashed
line.

Over the interval $35 \leq \omega _{v} \leq 55$~{\kms}, there are four
absorbers that lack the higher ionization phase typical for their
{\MgII} kinematic spread; they are ``{\CIV} deficient''.
These systems, Q~0117+213 at $z=1.0479$, Q~1317+277 at $z=0.6606$, 
Q~0117+213 at $z=0.7290$, and Q~1329+274 at $z=0.8936$ (in order of
decreasing $\omega _{v}$), lie $3.3$, $2.9$ $2.5$, and $2.3\sigma$
from the correlation line, respectively.
As compared to other DLAs with similar $\omega _{v}$, the DLA at
$z=0.6561$ in the field of Q~$1622+238$ (3C~336) also appears to have
a slight {\CIV} deficiency. 

\section{Discussion}

Three important observational facts are: 
(1) the scatter in $W_{r}(1548)$ vs.\ $W_{r}(2796)$ indicates that the
strength of {\CIV} absorption is not driven by the strong ``central''
{\MgII} component that dominates $W_{r}(2796)$,  
(2) $\omega_{v}$ is sensitive to the presence  of small $W_{r}(2796)$,
outlying velocity clouds, and
(3) the scatter of $W_{r}(2796)$ vs.\ $\omega_{v}$ is large, whereas
the scatter of $W_{r}(1548)$ vs.\ $\omega_{v}$ about the correlation
line is only $0.22$~{\AA}.
These facts imply that, independent of the overall {\MgII}
line--of--sight  kinematics, the existence and global dynamics of
smaller, kinematic ``outliers'' are intimately linked to the
presence and physical conditions of a higher ionization phase.  

It would appear that {\CIV} is governed by the same physical processes
that give rise to kinematic outlying {\MgII} clouds.
The {\CIV} absorption could arise due to the ionization balance 
in the {\MgII} clouds, ionization structure in the clouds, or 
due to multiphase structure.
In most cases, multiphase structure is the likely explanation
because the small $W_{r}(2796)$, single phase, outlying velocity clouds with
$b\simeq 6$~{\kms} cannot produce the large observed  $W_{r}(1548)$,
as shown by Churchill \etal (1999a\nocite{weak}) [also see Churchill \&
Charlton (1999\nocite{q1206})].

If the {\MgII} clouds are accreted by infall and/or minor mergers, 
such that the energetics originated gravitationally (e.g.\
\cite{mo96}), multiphase structure could arise from shock heating
and/or merger induced star formation (e.g.\ \cite{hernquist95}).
If, on the other hand, the absorbing gas is mechanically produced by
winds from massive stars, OB associations, or from galactic fountains
and chimneys, a dynamic multiphase structure could arise from shock
heated ascending material that forms a high ionization layer (corona)
and supports a descending lower ionization layer, which then breaks
into cool, infalling clouds [see Avillez (1999\nocite{avillez}), and
references therein].
Both scenarios imply a link between galaxy star formation histories,
in particular multiple episodes of elevated star formation [but not
necessarily bursting (\cite{dahlem98})], and the presence of outlying
velocity {\MgII} clouds and strong {\CIV}.

Infall models predict increasing cloud densities with decreasing
galactocentric distance (e.g.\ \cite{mo96}), resulting in an
ionization gradient (clouds further out are more highly ionized).
In Figure~\ref{fig:fig3}$a$, we show $W_{r}(1548)/W_{r}(2796)$
vs.\ impact parameter for 13 absorbing galaxies.
There is no \pagebreak obvious evidence for an ionization gradient (95\%
confidence). 
However, $W_{r}(1548)/W_{r}(2796)$ could be sensitive to halo mass
(e.g. \cite{mo96}), to the presence of satellite galaxies
(\cite{york86}), or to the sampling of discrete clouds over a range of
galactocentric distances along the line of sight.
In Figure~\ref{fig:fig3}$b$, we show $W_{r}(1548)/W_{r}(2796)$
vs.\ galaxy $B-K$ color for 11 galaxies.
There is a suggested trend ($2\sigma$) for red galaxies (those
dominated by late--type stellar populations) to have small
$W_{r}(1548)/W_{r}(2796)$.
If such a trend is confirmed in a larger data sample, it would not be
incompatable with a dynamical multiphase scenario in which absorbing
gas properties are linked to the host galaxy stellar populations, and
therefore, star formation history.

The tight correlation between {\MgII} kinematics and {\CIV} absorption
may imply a self--regulating process involving both ionization
conditions and kinematics in the halos of higher redshift, $\sim
L^{\ast}$ galaxies (e.g. \cite{lb92}), as explored by Norman \&
Ikeuchi (1989\nocite{norman}) and Li \& Ikeuchi (1992\nocite{li}).
Perhaps outflow energetics from supernovae during periods of elevated
star formation are balanced by the galactic gravitational potential
well, resulting in a fairly narrow range of kinematic and
multiphase ionization conditions.
Such a balance might set up a high ionization Galactic--like
``corona'' (\cite{savage97}) in proportion to the kinematics of
gravitationally bound, cooling material. 
All this would imply that galaxy ``coronae'' have been in place since 
$z \sim 1$, that their nature primarily depends upon the host
galaxy's star formation history, and therefore morphology,
environment, and stellar populations (as seen locally, e.g.\
\cite{dahlem98}).
Detailed observations of the stellar content and galactic morphologies
and space--based UV high--resolution spectroscopy of high ionization
absorption lines, would be central to establishing the interactive
cycles between stars and gas in higher redshift galaxies.

\vglue 0.1in
Support for this work was provided by the NSF (AST--9617185), and NASA
(NAG 5--6399 and AR--07983.01--96A) the latter from the Space
Telescope Science Institute, which is operated by ARUA, Inc.,
under NASA contract NAS5--26555
BTJ acknowledges support from NOAO, which is operated by ARUA,
Inc., under cooperative agreement with the NSF.

\end{document}